# A model for multi-attack classification to improve intrusion detection performance using deep learning approaches

Arun Kumar Silivery [a], Ram Mohan Rao Kovvur [b], Ramana Solleti [c], LK Suresh Kumar [a], Bhukya Madhu [d,*]

[a] *Department of CSE, University College of Engineering(A), Osmania University, Hyderabad, India*
[b] *Department of Information Technology, Vasavi College of Engineering, Hyderabad, India*
[c] *Department of Computer Science, Bhavans Vivekananda College, Hyderabad, India*
[d] *Department of CSE, KG Reddy College of Engineering & Technology, Hyderabad, India*

ABSTRACT

This proposed model introduces novel deep learning methodologies. The objective here is to create a reliable intrusion detection mechanism to help identify malicious attacks. Deep learning based solution framework is developed consisting of three approaches. The first approach is Long-Short Term Memory Recurrent Neural Network (LSTM-RNN) with seven optimizer functions such as adamax, SGD, adagrad, adam, RMSprop, nadam and adadelta. The model is evaluated on NSL-KDD dataset and classified multi attack classification. The model has outperformed with adamax optimizer in terms of accuracy, detection rate and low false alarm rate. The results of LSTM-RNN with adamax optimizer is compared with existing shallow machine and deep learning models in terms of accuracy, detection rate and low false alarm rate. The multi model methodology consisting of Recurrent Neural Network (RNN), Long-Short Term Memory Recurrent Neural Network (LSTM-RNN), and Deep Neural Network (DNN). The multi models are evaluated on bench mark datasets such as KDD'99, NSL-KDD, and UNSWNB15 datasets. The models self-learnt the features and classifies the attack classes as multi-attack classification. The models RNN, and LSTM-RNN provide considerable performance compared to other existing methods on KDD'99 and NSL-KDD dataset.

## 1. Introduction

In the present scenario, significant advancement of computer systems completely changed our daily lives and made our existence dependant on them. From small companies to large enterprises, individuals to government agencies, many of their activities are performed through network services [1]. With the tremendous growth of the use of Internet, our computer systems are exposed to elevate high amount of threats. Any vulnerability in the network devices and computing platforms can expose the network system under various attacks and may lead to catastrophic consequences [2].

While dealing with monitoring, prevention, detection, and reaction to cyber security events and prospective network attacks/intrusions, organisations should concentrate on cutting-edge technology to support human analysts [3]. It is quite difficult to successfully identify and stop the invasions in their tracks. Traditional network intrusion detection systems (NIDS) are rule- or signature-based systems that have not been sufficient for the rapidly expanding network and are unable to handle assaults because to their increasing volume, complexity, and deflation [4].

The firewall is designed to prevent unauthorised access to the whole network or systems. It has been shown that the firewall and its versions are readily circumvented by attackers, for instance by utilising a bogus source address. Moreover, a great deal of assaults like DoS and DDoS went undetected. An innovative security mechanism known as the "Intrusion Detection System" has been developed to address the shortcomings of current conventional security methods (IDS) [5].

The process of analyzing a computer system or network's activity and examining it for indications of potential events, such as violations or suspicious behavior of computer security regulations, acceptable usage guidelines, or standard security procedures, is known as intrusion detection [6]. There are too many possible causes of events, including dangerous malware (like worms and spyware), internet hackers who get remote access to networks, and privileged persons who misuse their privileges or attempt to maximize their authority for purposes for which they are not authorized. Intrusion detection systems are programs that maintain the intrusion detection technique (IDS) [7]. Identification of potential events is the IDS's main objective. In other words, IDS may be able to spot instances in which an attacker effectively exploited a system flaw. Many IDS may be configured with a set of firewall rules, such as settings, to aid in identifying network traffic that transgresses the company's security or permissible use policies. IDSs with this capability can watch data transfers and detect odd ones, such transferring a large database to a user's machine [8].

### 1.1. Intrusion detection methodologies

A variety of methods are used by IDS technology to detect intrusions. Anomaly-based detection and intrusion detection based on signatures are the two primary categories of detection technologies. The majority of IDS systems use a range of detection methods, either alone or in combination, to accurately identify intrusions [9].

#### 1.1.1. Signature-based detection

Signature-based detection, also known as misuse-based detection, is used to keep the database of signatures from recognized attacks current. A risk may be recognized by its signature, which is a pattern. If there is a match between the data gathered from the audit unit and the database, alarms can go out. These patterns represent suspicious collections of potentially risky practices or behaviors that have been recorded in databases. This system's primary benefit is how rapidly it can identify patterns and fingerprints and, after some acclimation, how well it can comprehend network dynamics. Attacks that have been effective in the past and whose patterns have been noted in the database should be kept up [10].

*1.1.1.1. Anomaly-based detection.* Anomaly-based IDS, also known as behavior-based IDS, is the second approach type. These systems learn the regular behavior of people, hosts, network connections, or applications rather than preserving the signatures of known assaults. The development of the behavior profiles involves monitoring the traits of typical behavior over time. A few behavioral traits that may be used to create profiles include a user's use of email, unsuccessful login attempts, and processor usage over a specific time frame [11]. As an intrusion detection system creates the initial profiles for its detection algorithms, it may spot malicious behavior. Every difference from the norm is regarded with suspicion, and a warning is given. These systems work under the assumption that any unusual activity or behavior deviates greatly from the norm. These systems are adept at identifying zero-day attacks by design [12].

*1.1.1.2. Hybrid based detection\.* Combining the signature and anomaly based techniques will increase the benefits and reduce the drawbacks of both approaches. By using this approach, it is possible to reduce the number of false alarms while increasing the rate of zero-day attack detection. No system, according to a study by A. Buczak, was solely based on signatures or anomalies; instead, IDSs are often installed as a combination of both signature and anomaly setup [13].

### 1.2. Problem statement

The primary focus is anomaly-based intrusion detection systems, and it significantly advances the field of such IDS. The high incidence of false alerts that anomaly-based IDS experiences is widely recognized. The high false positive rate is still being worked on. We think the issue may be treated as a classification issue since data analysis is involved in intrusion detection. This discovery further demonstrates the effectiveness of categorization systems since cleaner input data increases the likelihood of accurate outcomes. From an anomaly-based approach, it suggests that the false positive rate may be significantly reduced if we are successful in extracting characteristics that effectively distinguish normal data from abnormal data. So, in this study, we examine the methods for differentiating between normal and aberrant data [14].

Similar to the preceding observation, we see that the bulk of intrusion detection methods based on data mining and machine learning use well-established tools and procedures. These recommended techniques may end up being unsuccessful in precisely identifying data as normal or abnormal. As it is required to modify such strategies to meet the requirements of intrusion detection, we utilize deep learning and concentrate on this area of our work.

### 1.3. Artificial intelligence

Making robots behave like a human brain is the broad field of computer science known as artificial intelligence (AI). Using traditional computational approaches, it is utilised to elucidate complicated issues. The phrase "artificial intelligence" was first used in 1956 by computer scientists who wanted to know whether robots could "think" like people. Marvin Minksy described it as "the science of having robots accomplish things that would need intelligence if they were done by humans" [15]. Chollet's definition of artificial intelligence as "the endeavour to automate intellectual processes ordinarily done by humans" is another comparable concept. In order to achieve the objective of automating intelligent tasks that are typically performed by people, artificial intelligence encompasses various methodologies in addition to the subfields of artificial learning and deep learning. These methods work well for tackling clearly stated logical problems, such as those involved in games, but they are not well suited for dealing with more challenging tasks, such picture categorization and language translation. As a result, machine learning, a more recent approach to artificial intelligence, has surpassed other approaches to symbolic AI in popularity [16].

### 1.4. Machine learning

The study and development of systems from data are topics covered by a branch of artificial intelligence known as "machine learning" [17]. Machine Learning algorithms that adapt and gain knowledge from data are known as machine learning algorithms. Instead of actively instructing the computer and telling it what to do, machine learning approaches allow the software to indirectly learn what output to create depending on incoming data. This enables the computer to do tasks and make choices on a variety of data kinds that it has never seen before owing to learning from data.

### 1.5. Deep learning

Deep Learning, a complex kind of machine learning, employs many levels of data abstraction at different processing stages [18,19]. With the use of back-propagation, Deep Learning shows how a computer can learn the complex dataset structures while also adjusting internal settings at each layer. It succeeds in generalizing new instances when the data is complex and has a high degree of dimensionality. Deep learning also makes it possible to train nonlinear models on huge datasets.

Although deep learning has lately become more well-known, it has been present since the 1940s. The first algorithms for deep learning mainly took inspiration from biologically inspired computer models of the human brain. While it is still recognized as a major source of inspiration for deep learning, the neuro-scientific method is no longer the core pillar of the field. There is just a lack of knowledge about the basic mechanisms and algorithms the brain employs to operate. This is a current and active area of research within the field of "computational neuroscience" [20–25].

*1.6. Recurrent neural network (RNN)*

The RNN is a development over the conventional feed-forward neural network. The foundation of RNN is sequential data. Each input and each output in a traditional neural network, in our opinion, are distinctive. Since each element in a sequence is processed in the same manner, with the outcome depending on past computations, recurrent neural networks (RNNs) are so named. To get contextual information, each unit in an RNN receives both the past and current states [26,27].

*1.7. Long Short Term Memory recurrent neural network (LSTM RNN)*

As it develops over time, a deep neural network creates a FNN for each time-step. The gradient algorithm is then used to adjust the weights and biases for each hidden layer, minimizing the discrepancy between the expected and actual outputs. If the time steps are more than 5 to 10, the typical RNNs are unable to perform as intended. Long-term back propagation allows erroneous signals to erupt or disappear, resulting in variable weights that lower the network's efficacy. The problem of vanishing/exploding gradient in RNN is overcome using Gated Recurrent Units and Long Short Term Memory (LSTM) networks (GRUs). The LSTM employs a gating system to control long-term dependency [28–30].

Every time a step is taken, the cell state of the LSTM is sent along. The information going through is optimised via a gating technique. Input gate (it), forget gate (ft), and output gate are the three gates present in each LSTM cell (Ot). The contribution of the current input and the prior output to the new cell state is determined by the forget and input gate (ct). How much of ct is revealed as the input is determined by the output gate.

**2. Literature review**

The researchers have created many intrusion detection systems during the last ten years. They were created for host-based and network-based intrusion detection systems, respectively. The systems that are being offered are hybrid systems that combine HIDS/NIDS with signature- and anomaly-based technologies. The intrusion detection system is a kind of computational intelligence intelligence system. Computational intelligence's primary objective is to provide answers to challenging real-world issues. In order to build a more reliable IDS, a successful IDS must mix soft computing technologies with more complex mathematical procedures and traditional analytic methods. In this paper, we examine three crucial research areas with major implications for the suggested framework. Initially, both machine learning and deep learning methods for shallow learning intrusion detection are reviewed. We discuss current hybrid intrusion detection systems in the second part. Lastly, we provide some analyses and conclusions from the literature reviews [32].

For intrusion detection, deep learning, a branch of machine learning that has lately gained prominence, has been applied. Most studies have proven that deep learning completely outperforms traditional methods. The literature overview of the single- and hybrid-deep learning methodologies that have been used by researchers is covered in this part [39].

In order to tackle intrusion detection issues that are impacted by large dimensionality, high related characteristics, and high network data quantity, Levent Koc et al. [35] created a Hidden nave Bayes data mining approach. To improve the performance of the developed approach on the KDD99 dataset, they applied two well-known variation methods in this experiment: entropy minimization discretization and proportional k-interval discretization. The model contains a feature selection model based on three filter approaches, including INTERACT feature selection methods, correlation-based consistency-based, and filtering based on correlation. These methods aid in the suggested method's successful completion of the KDD'99 dataset. In terms of detection accuracy, error rate, and misclassification cost, the model exhibits generally improved results.

The FCANN technique, developed by Gang Wang et al. [36] and based on fuzzy clustering and artificial neural networks, is a revolutionary approach. To ensure homogeneity within the clusters and heterogeneity across the clusters, the provided data is divided into clusters. Every training subset's size and complexity are decreased by the fuzzy clustering module, which also increases the efficacy and efficiency of the following ANN module. Implemented ANN as traditional feed-forward neural networks trained with the back propagation technique to anticipate incursion. The fuzzy aggregation module is used to combine several NN results and minimise detection mistakes. In terms of accuracy and detection stability, the experimental findings utilising the KDD'99 dataset show efficacy, notably in low frequency attack detection, such as R2L and U2R assaults.

The hybrid intrusion detection approach described by Gisung Kim et al. [20] blends the misuse detection model and anomaly detection model's data decomposition structure hierarchically. Regular training data are divided into manageable subsets using the C 4.5 decision algorithm. In each deconstructed area, an anomaly detection model was created using the one-class support vector machine (1-class SVM). For the NSL-KDD dataset, the suggested approach is assessed. In this method, the test samples are categorised in a dissected area using one of the SVM models with a single class. The decomposition method might reduce the amount of time spent testing for abnormalities. The hybrid model reduced the train and test times while enhancing IDS performance.

A unique combined principal component analysis (PCA) and Support vector machine (SVM) was created by Sumaiya Thaseen et al. [21] by enhancing the kernel parameters using an automated parameter selection approach. By taking into account minority attacks like U2R and R2L, the KDD dataset is painstakingly split into train and test datasets. Early on, PCA chooses an appropriate subset of each characteristic while excluding those that are redundant or unneeded. To preserve a higher value that may be used to build the feature vectors, PCA often utilises a variance threshold. The necessary parameters are obtained from PCA in the second step as a train and test set for SVM to carry out a classification. In order to decrease train and test overhead time, the classifiers categorised the assaults using a smaller feature set.

Based on the intersection concept of mathematics, Ketan Sanjay and colleagues [22] suggested a feature selection strategy. Three distinct kinds of feature selection strategies are used in this method. application of the Genetic Algorithm (GA) with the correlation-based feature selection (CFS) technique. After CFS, intersection of various population sizes was obtained using the GA approach. With a pre-processed dataset, Information Gain (IG) and Correlation Attribute Evaluator (CAE), two additional filter techniques, are used with ranker as the search algorithm. J48 and Naive Bayes classifiers were used to categorize the testing characteristics. The accuracy, runtime, and chosen number aspects of the results are assessed using the NSL-KDD dataset.

Principle component analysis (PCA) and Support Vector Machine (SVM) were used to create Sumaiya Thaseen Ikram et alsuggested.'s hybrid intrusion detection model [23] To achieve the best feature subset with the greatest eigen values kept in each eigen vector, the PCA created eigen vectors for each feature. The class label is predicted using the SVM classifier. During the training step, the feature matrix is used to identify the class labels. To recognize the pattern of unknown traffic, the testing step uses the learning rules from the training phase. SVM classifier chooses automated parameters by optimizing kernel parameters. This technique reduces the train and test times while improving accuracy by optimizing the kernel parameter () and punishment factor (C). On the NSL-KDD dataset, the model's evaluation yields results.

Sasanka Potluri et al. [24] developed the Accelerated Deep Neural Network (DNN) to find abnormalities in network data. Its architecture consists of pre-training and fine-tuning techniques. In the pre-training process, two autoencoders are used. Out of the 41 features in the NSL-KDD dataset, one autoencoder with 20 neurons selected 20 features, while a second autoencoder with 10 neurons selected 10 features.

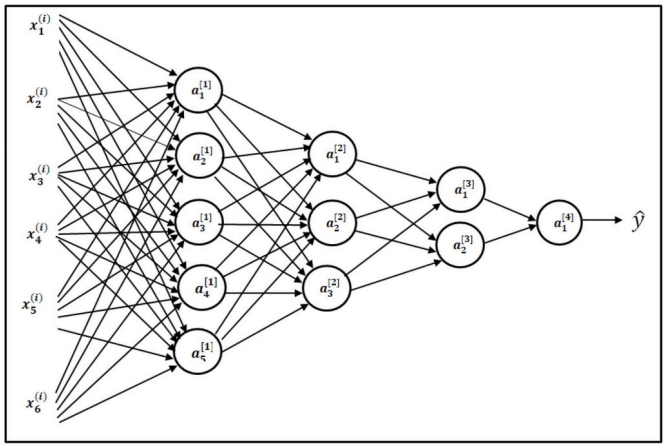

**Fig. 1.** Comprehensive neural network representation.

Phase one of fine-tuning employs a softmax classifier to classify the multiclass attacks after the pre-training approach. In fine-tuning step 2, the whole produced DNN is further adjusted via back propagation on all hidden layers. Its directed back propagation enhances the properties in the intermediate layers. As a result, the intrusion detection issue is better suited for the model.

Jihyun Kim et al. [25] develop an IDS model based on deep learning known as the Long Short Term Memory recurrent neural network using the KDDCup'99 dataset (LSTM-RNN). The model was trained and tested using 10% of the data. In the data ratio of the assaults, the DoS attack is given greater weight. Despite having a slightly greater FAR than the competition, the model performed the best in terms of DR and accuracy.

Thi-Thu-Huong Le et al. [26] found a suitable optimizer for the Long Short Term Memory recurrent neural network. The model is constructed using six optimizers: RMSprop, adagrad, adadelta, adam, adamax, and nadam. The model performance is assessed on the KDDCup'99 dataset in two circumstances. In Case 1, the average % classification of attack types is computed. To evaluate categorization performance, Case 2 computes metrics such as accuracy, detection rate, false alarm rate, recall, and efficiency. In both cases, the model performs better with a learning rate of 0.002.

Chuanlong Yin et al. [27] proposed Recurrent Neural Network IDS (RNNIDS) for binary and multi classification. The approach has two parts: backward propagation and forward propagation. Forward propagation is in charge of calculating the output values, while back propagation is in charge of communicating the residuals that collected to update the weights. The model evaluated the results using the NSL-KDD dataset. The effectiveness of the method was assessed in contrast to other well-known classification models including J48, Naive Bayesian, and Random Forest in terms of accuracy, detection rate, and false positive rate.

Nathan Shone et al. [28] proposed the nonsymmetric deep autoencoder for unsupervised feature learning (NDAE). The term "NDAE" refers to an autoencoder having numerous non-symmetric hidden layers. In essence, the proposed model does not use the symmetric encoder-decoder paradigm but just the encoder phase (non-symmetric). This is because decreasing computation time may have minimal impact on accuracy and output. With high-dimensional input, NDAE performs well as a scalable hierarchical unsupervised feature extractor. The NDAE uses the random forest to classify the feature learned data (RF). The model is evaluated with high levels of accuracy, precision, and recall coupled with a brief training time for the KDD'99 and NSL-KDD datasets.

Md Moin Uddin Chowdhury et al. [29] developed a few-shot deep learning technique to detect minority attack types in the KDD'99 and NSL-KDD dataset. The first training of Deep Convolutional Neural Network (CNN) is to recognize intrusions. The pre-processed train and test data are sent to the CNN architecture. After retrieving the intermediate features, the SVM and k-NN are given the features as input. They considered one neighbor while using the k-NN classifier. The model produced state-of-the-art outcomes on the KDD and NSL-KDD datasets.

Tao Ma et al. [30] presented the Spectral Clustering and Deep Neural Network (SCDNN) technique, a novel approach. The dataset is initially divided into k subgroups using a cluster center, also known as a spectral cluster, based on sample similarity. The distance between data points in a test and train set is then determined using similarity traits. An input layer, two hidden layers, a softmax layer, and an output layer are the five layers that make up each DNN. The two hidden layers in this model separately learn features from clustered data, whereas the top layer is a five-dimensional output vector layer. The kth DNN is given, in turn, the training subset data from the kth cluster center created by the SC algorithm during the clustering phase. The names of the trained sub-DNN models are sub-DNN1 through sub-DNNk. The k sub-DNNs are given the k testing data subsets after the k training data subsets have been finished. Each sub-final DNN's output is combined, and the positive detection rates are evaluated. The model was evaluated using the NSL-KDD dataset and the KDDCup'99 dataset.

Shahriar Mohammadi et al. [31] introduced deep learning as a method for determining the finest qualities. The data is first normalized using the z-score. This method is characterized by a distribution with a mean of zero and a standard deviation. In the second stage, a deep learning autoencoder model is used to condense a large number of features. The final classification function of the algorithm, which can differentiate between normal and abnormal traffic, is based on the acquired features. The performance of the recommended model is evaluated on the KDD and NSL-KDD datasets.

Based on SVM and Multi-Layer Perceptron, Kazem Qazanfari et al. [32].'s hybrid anomaly-based intrusion detection method detects intrusions (MLP). Using a feature selection approach based on the entropy of the features is the first stage in extracting features from the KDD dataset. The data are then categorised using the state-of-the-art supervised learning methods SVM and MLP. Using a rule-based methodology, the results of two learning-based approaches are merged. The simulation results show that the recommended hybrid method is effective.

Divyatmika et al. [33] proposed a two-tier architecture to detect intrusions at the network level. The two categories of network behavior are abuse and anomaly detection. After pre-processing the data in the first stage, it uses hierarchical agglomerative clustering to build an autonomous model on the training set. Also, the dataset is classified using KNN. The MLP algorithm is used to identify abuse. An method called reinforcement is used to find abnormalities.

K. Shrivas et alANN-Bayesian.'s Net-GR method [34] is based on ensemble and feature selection methods. In this model, the artificial neural network and the Bayesian network are combined (ANN). Gain Ratio feature selection is used to remove superfluous features and increase classification precision. The ensemble model offers more accuracy than standalone models like ANN and Bayesian Net. The model is evaluated using both the NSL-KDD dataset and the KDD99 dataset.

Rana Aamir et al. [35] implemented a novel fuzziness-based semi-supervised learning approach. Unlabeled data are used in combination with a supervised learning strategy to increase classifier performance. Using a single hidden layer feed-forward neural network, a fuzzy membership vector and class classification are generated (SLFNN). The classifier maintains each category separately in the initial training set. Using the NSL-KDD dataset, the experiment is evaluated and contrasted with the existing classifier strategy.

A model for multi-class classification was built by Sumaiya T. I. et al. [36] using chi-square feature selection and SVM. A parameter tuning strategy is used to optimize the overfitting constant "C" and the gamma () kernel parameter of the Radial Basis Function. The multi class SVM reduces the train and test times. The experiment was examined using the NSL-KDD and KDD'99 datasets. As compared to other conventional techniques, it demonstrated a greater high detection rate and a lower

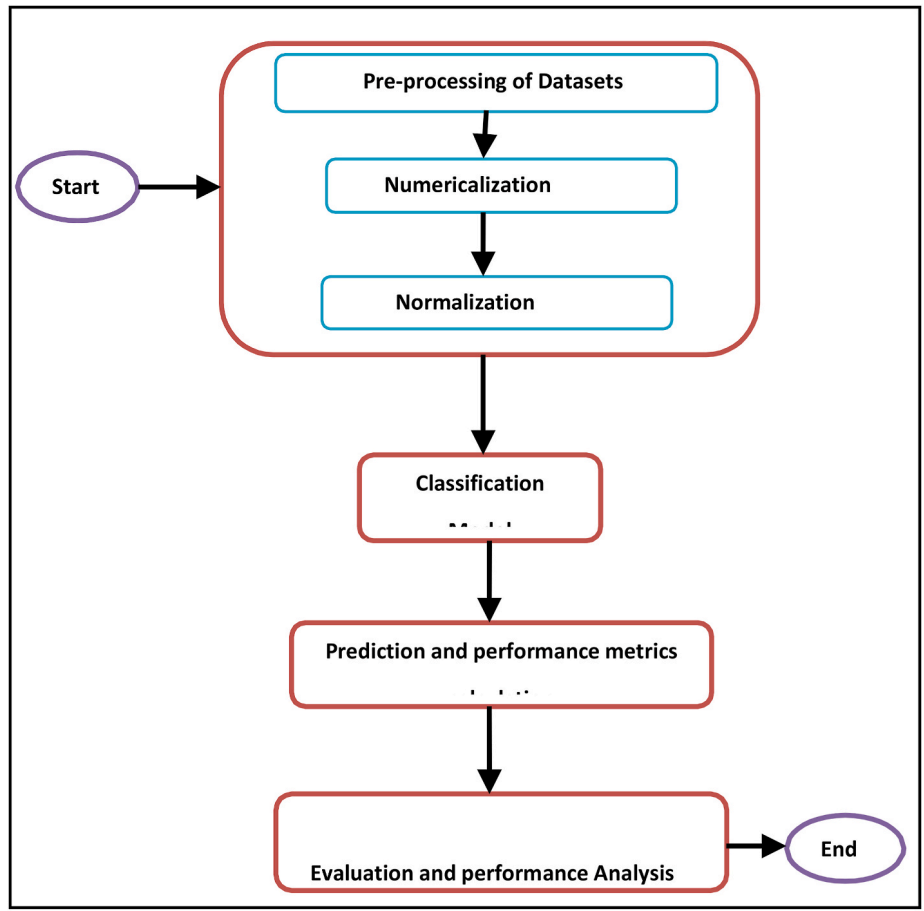

**Fig. 2.** Flowchart of proposed methodology.

**Table 1**
Distribution of NSL-KDD dataset sample distribution.

|  | Normal | DoS | Probe | U2R | R2L | Total |
|---|---|---|---|---|---|---|
| **Train** | 97,277 | 3,91,458 | 4107 | 1126 | 52 | 4,94,020 |
| % | 19.69 | 79.24 | 0.83 | 0.01 | 0.23 | 100 |
| **Test** | 9711 | 7458 | 2421 | 200 | 2654 | 22,544 |
| % | 43 | 33 | 11 | 0.9 | 12.1 | 100 |

false alarm rate.

Subba et al. [37] proposed a three-layer Artificial Neural Network (ANN) model including feed-forward and backpropagation techniques as well as other optimisation methods. It lessens the total computational load while maintaining good performance. The NSL-KDD dataset is used to evaluate the model's performance and compare it to that of other conventional models.

Min-Joo Kang et al. [38] created an effective IDS based on deep neural networks for the security of the in-vehicle network. The probability-based feature vectors are extracted from the network packets using unsupervised deep belief networks, and the conventional stochastic gradient descent method is then used. The system can recognize any malicious attack on the vehicle since the DNN provides the probability of each class to discriminate between regular and attack packets.

### 2.1. Research Gap

Let's talk about the numerous methods that various writers have used in their study efforts in this subject. The majority of studies employed classifiers and cluster algorithms as an intrusion detection system, according to the literature review stated above. Subsequently, coupled with classifiers, feature extraction techniques are employed to extract the crucial characteristics. Several researchers in this important field have used hybrid classifiers for intrusion detection, which include feature extraction, clustering, and classification techniques. Ensemble classifiers were employed as Intrusion Detection approaches by a very small number of researchers. The majority of the suggested methods were tested using the KDD CUP 1999 dataset, and more recently, the researchers utilised the NSL-KDD dataset, which is an updated version of

**Table 2**
The overall classification performance (%) of three models on KDD99, NSL-KDD, and UNSW-NB15 datasets.

| Dataset | KDD'99 | | | NSL-KDD | | | UNSW-NB15 | | |
|---|---|---|---|---|---|---|---|---|---|
| Methodology | RNN | LSTM | DNN | RNN | LSTM | DNN | RNN | LSTM | DNN |
| Accuracy | 98.73 | 96.85 | 98.20 | 98.68 | 96.25 | 98.36 | 77.45 | 80.39 | 80.39 |
| Detection Rate | 99.57 | 99.82 | 99.28 | 99.6 | 99.82 | 99.87 | 100 | 100 | 100 |
| Precision | 98.19 | 94.63 | 97.52 | 98.07 | 93.57 | 97.25 | 70.51 | 72.23 | 72.23 |
| F1-Score | 98.87 | 97.16 | 98.39 | 98.83 | 96.6 | 98.54 | 82.71 | 83.88 | 83.88 |
| FAR | 2.33 | 2.47 | 56.38 | 6.63 | 7.81 | 40.02 | 3.15 | 40.02 | 40.02 |

**Table 3**
Multi attack classification performance of accuracy (%) on KDD'99 and NSL-KDD datasets.

| Dataset | Model | DoS | Probe | R2L | U2R | Normal |
|---|---|---|---|---|---|---|
| **KDD'99** | RNN | 99.97 | 99.49 | 99.02 | 99.34 | 99.63 |
| | LSTM | 99.99 | 97.05 | 97.03 | 99.81 | 99.80 |
| | DNN | 99.99 | 98.75 | 98.41 | 99.64 | 99.60 |
| **NSL-KDD** | RNN | 100.00 | 99.82 | 98.84 | 98.92 | 99.77 |
| | LSTM | 100.00 | 96.63 | 96.34 | 99.62 | 99.90 |
| | DNN | 100.00 | 98.54 | 98.43 | 99.82 | 99.92 |

**Table 5**
Multi attack classification performance detection rate(%) on KDD'99 and NSL-KDD datasets.

| Dataset | Model | DoS | Probe | R2L | U2R | Normal |
|---|---|---|---|---|---|---|
| **KDD'99** | RNN | 100.00 | 95.45 | 96.01 | 53.73 | 99.70 |
| | LSTM | 99.98 | 97.64 | 78.85 | 71.64 | 99.77 |
| | DNN | 100.00 | 92.97 | 93.41 | 34.32 | 99.98 |
| **NSL-KDD** | RNN | 100.00 | 99.91 | 91.05 | 46.26 | 100.00 |
| | LSTM | 100.00 | 94.09 | 76.49 | 67.16 | 99.98 |
| | DNN | 100.00 | 93.30 | 93.51 | 71.64 | 100.00 |

the KDD'99 dataset. After thoroughly analysing the literature and approaches stated above, we came to the following conclusion:

✓ In recent years, numerous researchers have developed hybrid classifier algorithms using feature selection methods. Comparing these techniques to single classifiers, they achieved greater accuracy. Yet, there hasn't been much research done on ensemble classifiers.
✓ The methods for feature selection Principle Component Analysis (PCA), Correlation Feature Selection (CFS), and other methods were employed in machine learning to extract feature selection, and some researchers utilised autoencoders to do so. Higher performance was generated by the unsupervised autoencoders.
✓ The majority of the approaches made use of benchmark datasets like the NSL-KDD and KDD CUP 1999 datasets. Several academics have employed data mining and machine learning approaches to assess approximations of the actual performance of intrusion detection in real-time data. The vast majority of these methods were developed using shallow learning architectures. Some architectural designs still fall short in terms of intrusion detection.
✓ Several studies have categorised two classes as either an attack or as normal.
✓ Due to a lack of labelled training data and unpredictable network traffic, traditional machine learning algorithms are ineffective at detecting zero-day attacks and low frequency assaults like R2L and U2R. With a large number of datasets, deep learning algorithms are appropriate for feature learning and classification.

We get the conclusion that using a feature reduction technique is crucial to lowering the computing expense and improving classifier performance from the research results and review of the literature. It is challenging to create a single classifier approach due to the benefits of feature selection and feature extraction in detecting assaults using classifiers. It is advised to employ a hybrid strategy that combines feature extraction with feature selection to improve intrusion detection accuracy. As Deep Learning algorithms are more adaptive to systems with greater processing speeds and Graphics Processor Units, it is thereby advised for an enhancement of intrusion detection in the area of cyber security (GPU).

## 3. Proposed methodology

The proposed deep learning models are Recurrent neural network (RNN),Long Short Term Memory- Recurrent Neural Network (LSTM-RNN), and Deep Neural Network (DNN). These models consisting of pre-processing, feature extraction, training and testing phases. The raw data attributes consist of umerical and non-numerical values. Deep learning algorithms only required numerical attributes as an input. The numericalization process performs conversion of non-numerical attributes in to numerical attributes. The numericalization process can be done with label encoder. Label encoding assigns unique feature values to the non-numerical features. Some numericalized data attributes consist of large feature values and some feature values consist of minimum value. The difference between minimum and maximum feature value is very large. This difference is affected the original feature values. The normalization process avoids the effectiveness of the original feature values. Normalization could be done with min-max normalization. The normalized data is given as an input to our proposed hybrid methodology. The deep learning models self-learned the features and classify the data as multi attack classification. The classifier training and testing perform on train and test preprocessed datasets. The goal of this paper will be the following;

1) To implement a proposed deep learning models RNN, LSTM-RNN and DNN.
2) To train and test classifiers on real world benchmark datasets KDD'99, NSL-KDD and UNSW-NB15. ;
3) To provide obtained results. ;
4) To provide evidence of why this procedure can outperform the existing classification techniques. The experiment evaluation consists of the following phases.
   i. Selected datasets are pre-processed using the techniques numericalization and normalization. The numericalization used to transform categorical features in to numerical features. Normalization can be normalized the large feature set values between 0 and 1.
   ii. The proposed models RNN, LSTM, and DNN used to train the KDD'99, NSL-KDD and UNSW-NB15 datasets individually.
   iii. The models also used to test the datasets
   iv. The classifiers results evaluated and compared with existing classifiers results.

### 3.1. Recurrent neural network (RNN) and long short term Memory recurrent neural network

RNN is a variant of artificial neural network (ANN), and LSTM-RNN is improved version of RNN and that is a capable of order learning dependence in sequence prediction problems. The architecture of RNN and LSTM-RNN classifiers are discussed below.

#### 3.1.1. Architecture of deep neural network
Deep feed forwarded networks and multilayer perceptrons are two names for the deep neural network approach to deep learning (MLPs). It is one of the effective supervised learning techniques. DNNs may express functions with ever more complex functions by building neural networks with more layers and more units per layer [10]. Any network with more than two hidden layers is considered a deep neural network. A neural network is referred to as deep if it contains more than three layers, including the input and output layers.

**Table 4**
Multi attack classification performance accuracy (%) on UNSW-NB15 dataset.

| Dataset | Model | Exploits | Reconnaissance | Backdoor | DoS | Analysis | Fuzzers | Worms | Shellcode | Generic | Normal |
|---|---|---|---|---|---|---|---|---|---|---|---|
| **UNSW- NB15** | **RNN** | 99.99 | 99.99 | 99.88 | 99.46 | 99.56 | 99.93 | 99.73 | 91.00 | 70.05 | 78.84 |
| | **LSTM** | 100.00 | 99.99 | 98.65 | 97.92 | 98.16 | 98.84 | 99.72 | 99.18 | 83.85 | 84.43 |
| | **DNN** | 100.00 | 99.99 | 98.65 | 97.92 | 98.16 | 98.84 | 99.72 | 99.18 | 83.85 | 84.43 |

Fig. 1 depicts a typical deep neural network. It is a fully linked neural network since each input layer neuron is coupled to every neuron in every subsequent layer. The nodes in this diagram stand in for the inputs, and the edges for the weights or biases. The ith training example is indicated by the superscript I, while the lth layer is indicated by (l). Deep neural networks are appropriate for improving performance, particularly in the context of supervised learning, as a result of an increase in input data volume [3]. Fig. 1 may be used to illustrate the idea.

The model mathematically defined as $O: \mathbb{R}m \times \mathbb{R}n$. The input vector $x = x1,2,x3,.......,xm$ and its size is 'm' and the output vector is $O(x)$ and its size is 'n'. The computation of each hidden layer $hj$ defined mathematically as

$$h(x_j^{l+1}) = f(Z_{ij} + bj^{(l+1)}) \quad ... ... ... ... ... ...(1)$$

$$Z_{ij} = x_i^l w_{ij}^{(l,+1)} \quad ... ... ... ... ... ... ....(2)$$

All the lower layer neurons those are connected to neuron $j$. In equation 1 and 2 $x_i^{(l)}$ is neuron i activation function at layer $l$ and $Z_{ij}$ is the contribution of neuron $i$ at layer $l$ to the activation of neuron $j$ at layer $l+1$. The function $f$ is the non-linear activation function, $w_{ij}^{(l,+1)}$ is the weight and $bj^{l+1}$ is bais of neuron $j$.

*3.1.2. KDD'99 dataset description*

KDD'99 has been the dataset that is most often used since 1999 to evaluate anomaly detection systems. Based on data gathered from the DARPA's 1998 IDS evaluation study, it was developed by Stolfo et al. [5]. DARPA'98 consists of around 5 million connection records with a total of about 100 bytes each, which may be divided into approximately 4 gigabytes of compressed raw (binary) tcpdump data from 7 weeks of network activity. Over 2 million connection records are included in the test data, which spans two weeks. The KDD training dataset consists of more than 4 lakh 90,000 records in a single connection vector, each of which includes 4190 features and is categorised as either a regular record or an attack with exactly one specific attack type. These four categories encapsulate the simulated attacks the best.

*3.1.2.1. Dataset attack categories*

*3.1.2.1.1. DoS.* Denial of Services (DoS) exhausts the victim's resources, rendering it unable to respond to valid requests.

*3.1.2.1.2. Probing.* The probing attack gathers data about a target system's potential weaknesses that may then be utilised to conduct attacks against those systems.

*3.1.2.1.3. Remote to local (R2L).* Remote to local access refers to the capacity to dump data packets to a remote system across a network and get access as a user or root to do undesirable operations (R2L).

**User to Remote (U2R):** Attackers employ security weaknesses to get access to the system while logging in as ordinary users. The test data differs from the training data's probability distribution and includes several attack types that weren't included in the training data, which makes the challenge seem more realistic. Many new attacks are variations of well-known assaults, according to some experts, and the signature of well-known attacks may be used to identify novel versions. The dataset solely includes test data and consists of training assaults of 24 kinds and an additional 14 types.

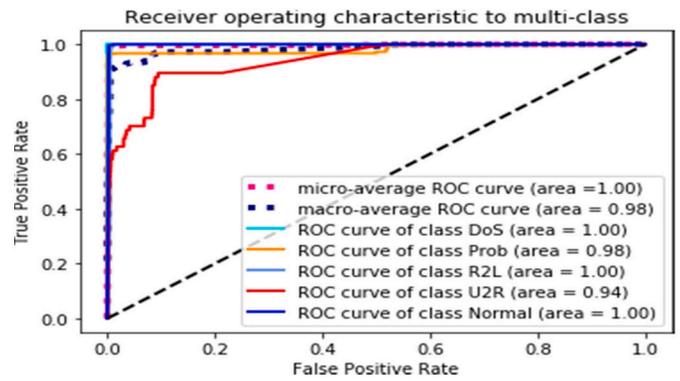

**Fig. 3.** The ROC curve and AUC of RNN model on KDD'99 dataset.

There are training dataset contain 4,94,020 records, and testing dataset contain 22,543 records. KDD'99 and NSL-KDD datasets have similar 41 fetures.

*3.1.3. NSL-KDD dataset description*

NSL-KDD is the refined version of KDD'99 dataset. It consists of 1,25,973 train records and 22,544 test records. NSL-KDD dataset features, feature groups, attack types and distribution of train and test dataset.

*3.1.3.1. Dataset selection.* We used benchmark datasets including KDD'99, NSL-KDD, and UNSW-NB15 datasets to develop the hybrid intrusion detection system. from those datasets kddcup99 csv that are accessible. For training and kdd test, csv was utilised. KDDTrain+'s testing csv from 1999. KDDTest+ and training utilize csv. Finally, UNSW NB15 training-set and NSL-KDD both utilised csv for testing purposes. UNSW NB15 testingset and training csv utilised. used for testing is csv.

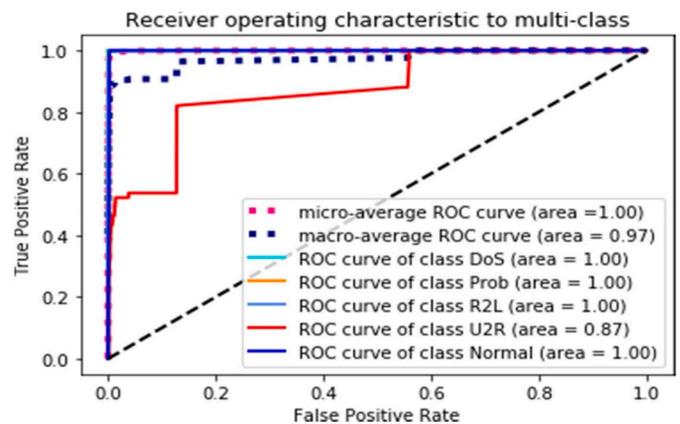

**Fig. 4.** The ROC curve and AUC of RNN model on NSL-KDD dataset.

**Table 6**
Individual attack classification performance detection rate(%) on UNSW-NB15 dataset.

| Dataset | Model | Exploits | Reconnaissance | Backdoor | DoS | Analysis | Fuzzers | Worms | Shellcode | Generic | Normal |
|---|---|---|---|---|---|---|---|---|---|---|---|
| **UNSW-NB15** | **RNN** | 100.00 | 99.97 | 98.97 | 97.43 | 43.27 | 99.91 | 6.81 | 50.79 | 98.89 | 52.38 |
| | **LSTM** | 100.00 | 100.00 | 99.31 | 73.09 | 10.48 | 85.03 | 0.00 | 51.58 | 97.42 | 65.37 |
| | **DNN** | 100.00 | 100.00 | 99.31 | 73.09 | 10.48 | 85.03 | 0.00 | 51.58 | 97.42 | 65.37 |

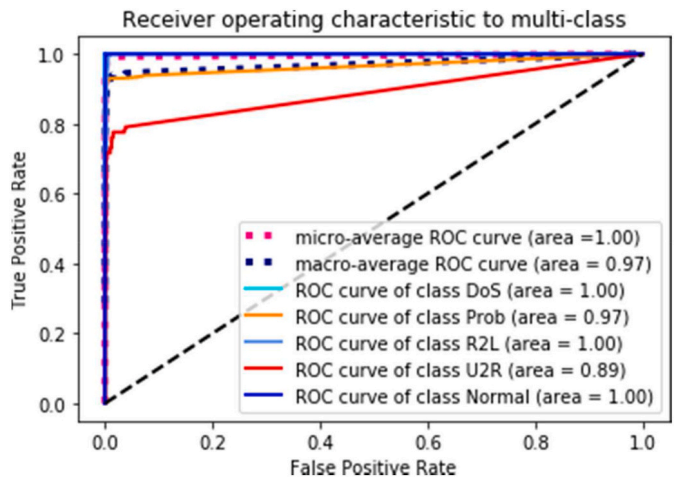

**Fig. 5.** The ROC curve and AUC of DNN model on NSL-KDD dataset.

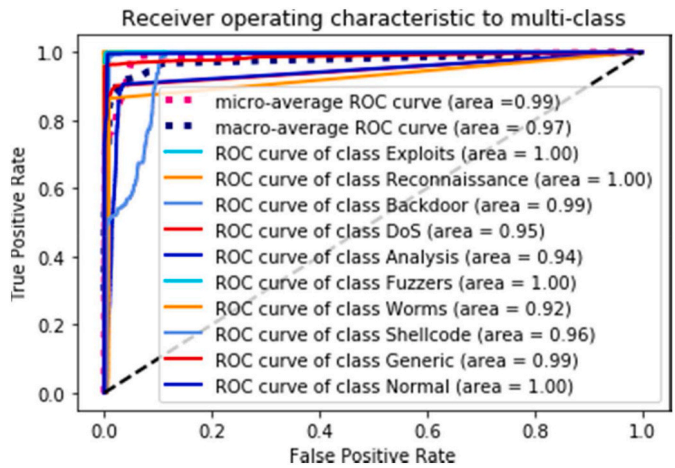

**Fig. 6.** The ROC curve and AUC of RNN model on UNSW-NB15 dataset.

## 3.2. Dataset pre processing

### 3.2.1. One-hot encoding

Labels are converted into numeric form using one-hot encoding so that machines can read them. The category value is changed to a numeric value that falls between 0 and the number of classes minus one. The choice of how to employ such labels may therefore be more successfully determined by learning algorithms. For the structured dataset, it is a key supervised learning pre-processing step.

### 3.2.2. Min-max normalization

When dealing with extremely wide differences in attribute values, normalization is often necessary. These significant differences in attributes might reduce performance effectiveness. Both of these feature settings are incompatible and inappropriate for use. A method called min-max normalization transforms the initial feature data using a linear transformation. The feature's minimum and maximum values are retrieved, and then each value is changed using the formula below.

$$x' = \frac{X - X_{min}}{X_{max} - X_{min}}$$

Where, accordingly, x is the source and $x'$ is a normalized set of data. Data normalization may quicken gradient descent learning to get the best outcome. According to the min-max normalization process, all attribute feature values are scaled from zero to one.

## 3.3. Proposed model

The end to end flow diagram of the implemented hybrid intrusion detection system is represented in the following Fig. 2. The datasets are preprocessed using numericalization and normalization. The preprocessed train and test data passed as input to the models. The models predict malicious and classify as multi class malicious then evaluated and analyze the performance.

### 3.3.1. Datasets used and pre-processing

The datasets KDD'99, NSL-KDD and UNSW-NB15 datasets are used to evaluate the performance of deep learning based hybrid intrusion detection system integrated Sparse autoencoder with deep neural network. At the preprocessing phase, the datasets features encoded using One-hot-encoding to transform categorical features into numerical features. The datasets features have been scaled using Min-Max normalization.

### 3.3.2. Tools and environment and technical details

In order to implement the hybrid intrusion detection model, the required tools and environment. The required environment is jupyter notebook, and packages are anaconda, pandas, numpy, scikit-learn, and Matplotlib.

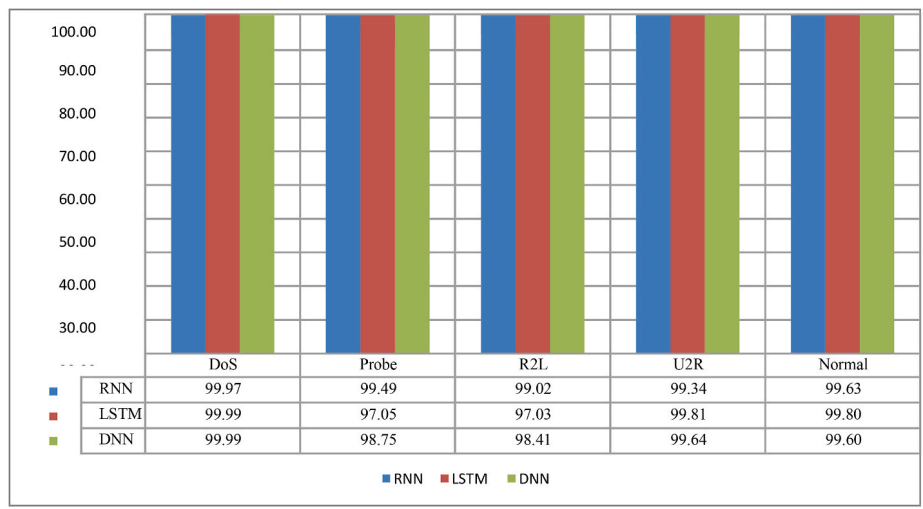

**Fig. 7.** Accuracy (%) comparison of classification methods on KDDCup99 dataset.

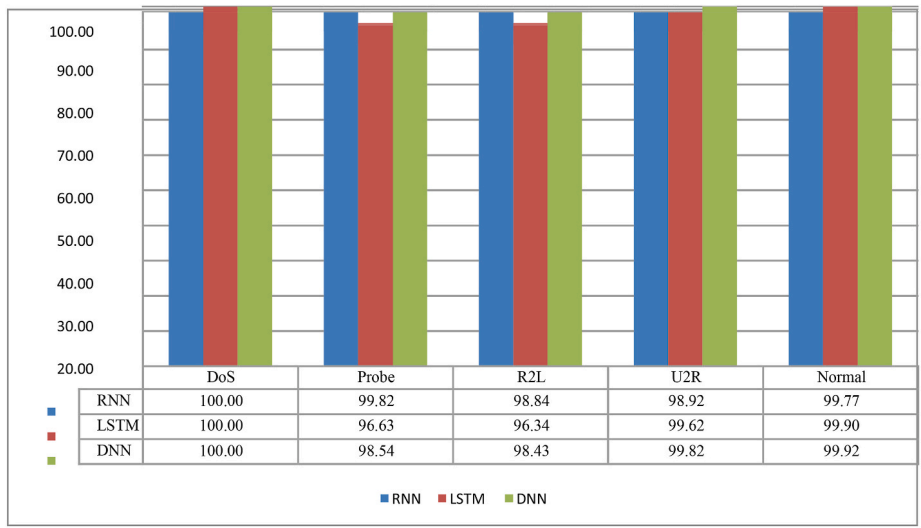

**Fig. 8.** Accuracy (%) comparison of classification methods on NSL-KDD dataset.

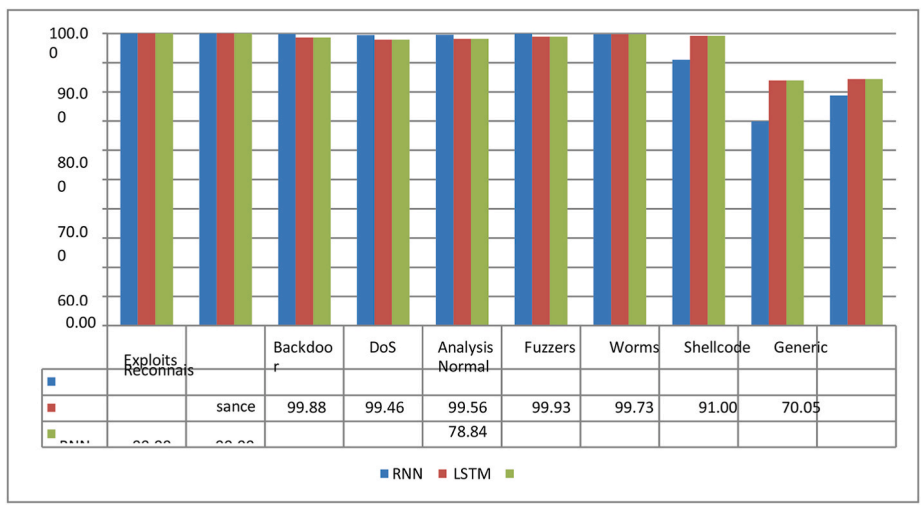

**Fig. 9.** Accuracy (%) comparison of classification methods on UNSW-NB15 dataset.

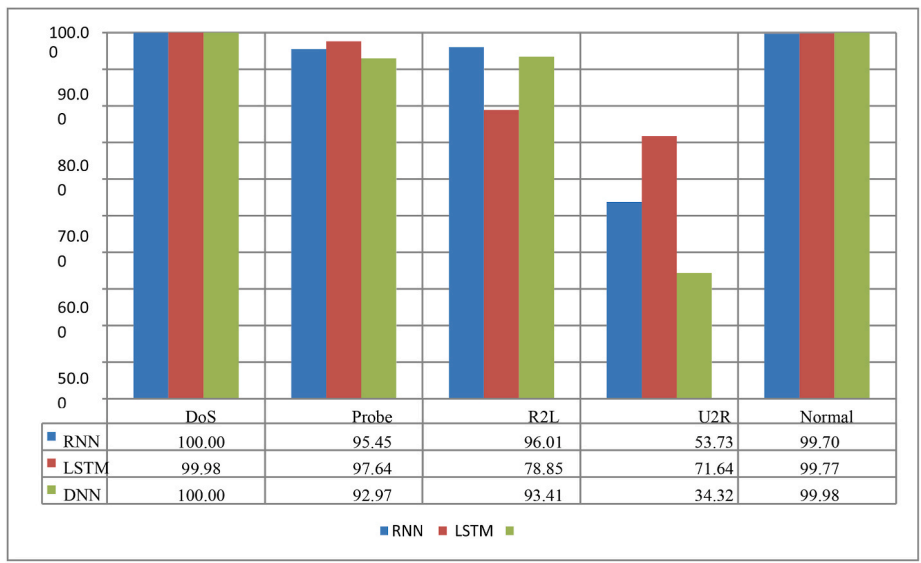

**Fig. 10.** Detection rate (%) comparison of various classification methods on KDDCup99 dataset.

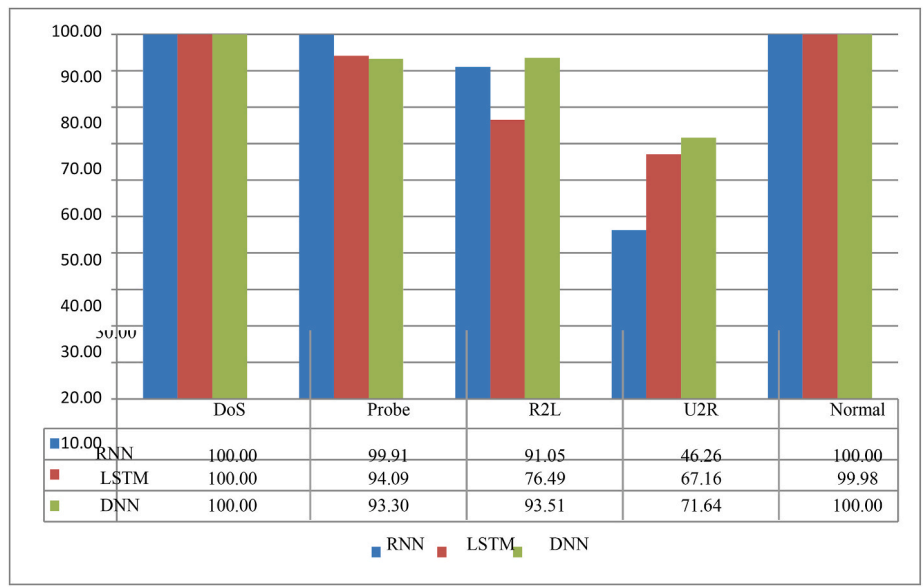

**Fig. 11.** Detection rate (%) comparison of various classification methods on NSL-KDD dataset.

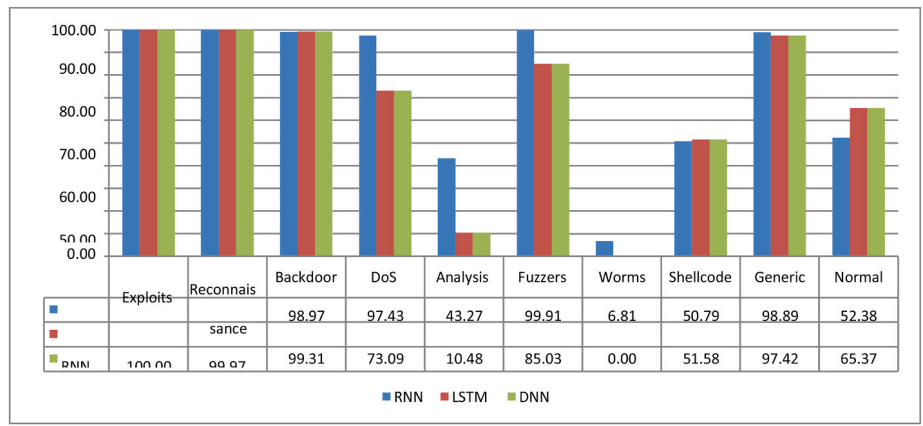

**Fig. 12.** Comparison of several classification algorithms' detection rates (%) using the UNSW-NB15 dataset.

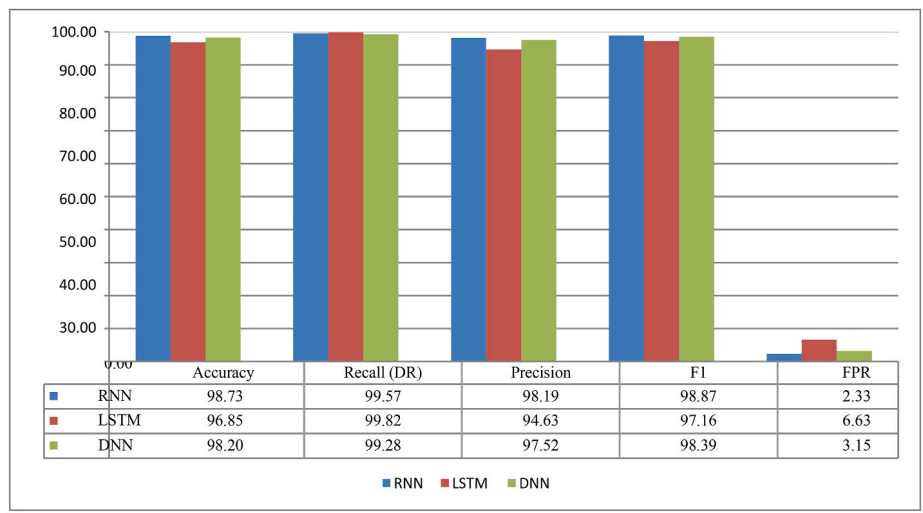

**Fig. 13.** Proposed models overall performance (%) comparison on KDD'99 dataset.

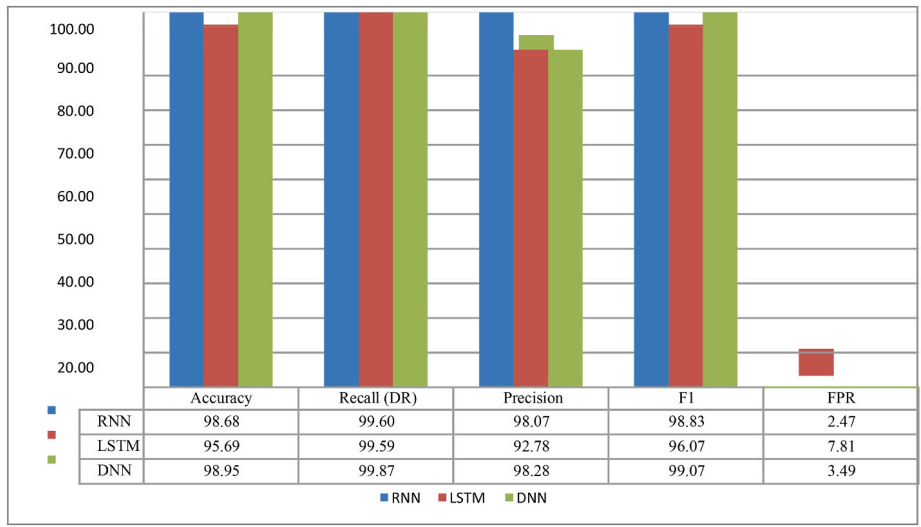

**Fig. 14.** Proposed models overall performance (%) comparison on NSL-KDD dataset.

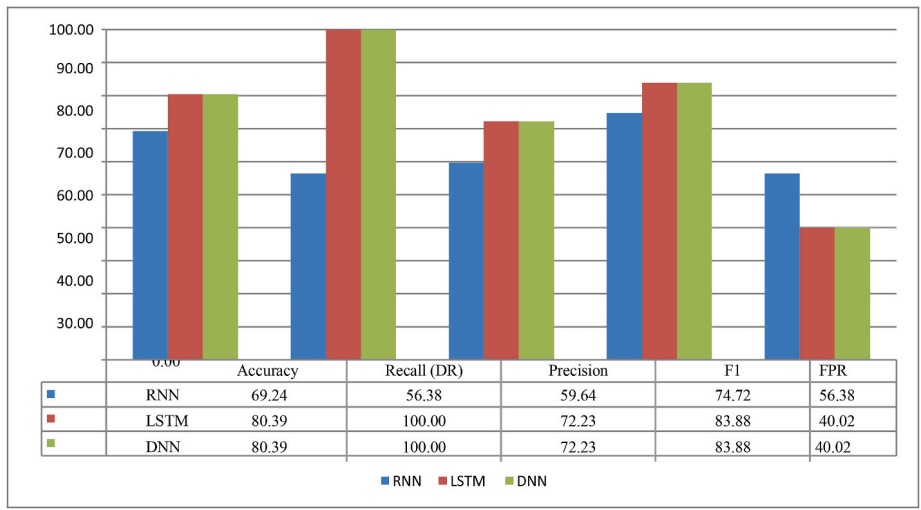

**Fig. 15.** Proposed models overall performance (%) comparison on UNSW-NB15 dataset.

Table 7
The comparison results of RNN and DNN with advanced methodologies on KDDCup99.

| Method | Accuracy | DR | F1 | FPR |
| --- | --- | --- | --- | --- |
| LSTM-RNN [124] | 96.93 | 98.88 | NA | 10.04 |
| RBM-DBN [125] | 97.16 | NA | NA | 0.48 |
| CNN-GRU [126] | 98.1 | 97.6 | 98.8 | NA |
| Multi-scale CNN [127] | 94.11 | 93.21 | NA | 2.18 |
| SAE [133] | 94.71 | NA | NA | 0.42 |
| DBN [134] | 93.49 | NA | NA | 0.76 |
| DNN [135] | 93.00 | NA | NA | 0.95 |
| SAE [136] | 97.85 | NA | NA | 2.15 |
| Our Proposed Model RNN | 98.73 | 99.57 | 98.87 | 2.33 |
| Our Proposed Model DNN | 98.20 | 99.28 | 98.39 | 3.15 |

Table 8
The comparison results of RNN and DNN with advanced methodologies on NSL-KDD dataset.

| Method | Accuracy | DR | F1 | FPR |
| --- | --- | --- | --- | --- |
| SAE-SVM [128] | 80.48 | NA | NA | NA |
| S-NDAE [129] | 85.42 | 85.42 | 87.37 | 14.58 |
| ID-CVAE [130] | 80.10 | 80.10 | 79.08 | 8.18 |
| SAVAER-DNN [126] | 89.36 | 95.98 | 90.08 | 4.70 |
| AE [136] | 89.92 | NA | NA | 10.78 |
| Ensemble [137] | 92.90 | NA | NA | 0.92 |
| **Our Proposed Model RNN** | **98.68** | **99.60** | **98.83** | **2.47** |
| **Our Proposed Model DNN** | **98.95** | **99.87** | **99.07** | **3.49** |

### 3.3.3. Methodology implementation

The performance of anomaly-based intrusion detection across the KDD'99, NSL-KDD, and UNSW-NB15 datasets was assessed using the hybrid deep learning intrusion detection system SAE-DNN. We implemented the RNN, LSTM, and DNN models as classification models using a three-layer, medium-sized architecture. After implementing basic models, hybrid models integrated SAE with models like RNN, LSTM, and DNN stated above. Scikit-learn 0.20.0 has been used to construct the basic and hybrid models. There are two methods to put the models into practice: either from scratch in programming, or utilising pre-written libraries like Keras or TensorFlow. It takes a lot of effort and technical knowledge to create neural network programs from scratch. As a result, we decide to employ Tensorflow, a deep learning framework.

The attack class type is also has categorical values. These categorical values also need to be convert into numerical. For binary classification, the numeric values '0' and '1' for attack and normal labels respectively. In the multi attack classification process for the dataset KDD'99 and NSL-KDD datasets attack types are numericalized as 0, 1, 2, 3, and 4 for DoS, Prob, R2L, U2R, and normal respectively. For the UNSW-NB15 dataset the attack types numericalized as 0, 1, 2, ….., and 9 for exploits, Reconnaissance, backdoor, DoS, analysis, fuzzers, worms, shellcode, generic and normal respectively.

We developed straightforward RNN, LSTM, and DNN models, as well as a hybrid model that combined the three with a Sparse autoencoder. The input layer is the first layer for basic models, the hidden layer is the second layer, and the output layer is the third layer. With the KDD'99 and NSL-KDD datasets, there are 41 neurons in the input layers, while 43 neurons are present in the UNSW-NB15 dataset. The output layer is made up of 5 neurons for the KDD'99 and NSL-KDD datasets and 10 neurons for the UNSW-NB15 dataset. The hidden layer is made up of 128 neuron units. Scikit-learn 0.20.0 was used to implement the models.

Using the datasets from KDD'99, NSL-KDD, and UNSW-NB15, we developed the RNN, LSTM, and DNN basic models. When these models were implemented, we created a hybrid model that combined RNN, LSTM, and DNN models with a sparse autoencoder to improve performance. On the datasets from KDD'99, NSL-KDD, and UNSW-NB15, these hybrid models are assessed.

### 3.3.4. Hyper-parameters

The hyper-parameters plays vital role to improve the results performance of learning algorithms. kalusGreff et al. performed analysis on the impact of hyper-parameters. The proposed single models RNN, LSTM and DNN are executed with the following hyper-parameters.

The classification models RNN, LSTM and DNN executed after pre-processing the train and test data. The datasets are pre-processed with help of above mentioned techniques such as one-hot-encoding and min-max normalization. The pre-processed train and test data given as a input to the individual dataset to the individual models. Every model executed with three datasets. After execution of the models we evaluated intrusion detection performance. The models performance evaluated in terms of detection rate, accuracy false alarm rate and F1 measure. The following section describes the performance of the models on three described datasets (see Table 1).

## 4. Results and discussion

To present the efficacy and performance of the proposed classifiers, we used publicly available intrusion detection train and test datasets. The following Table 2 depicts the performance of three models RNN. LSTM-RNN and DNN on KDD'99, NSL-KDD and UNSW-NB15 datasets.

From the analysis of the above results we investigated that the model RNN outperformed with the other models on KDD'99 and NSL-KDD dataset. The model DNN is outperformed with the other models on NSL-KDD and UNSWNB15 datasets. But the model LSTM-RNN shown significant results on KDD'99 and NSL-KDD dataset compared with existing shallow learning algorithms. The detailed results of multi-class attack classification of three models are described in the following Tables 3 and 4. The Table 3 shows the detection rate of the three classifiers on KDD'99 and NSL-KDD dataset and Table 4 shows the detection rate of the UNSW-NB15.

The Table 5 shows the detection rate of the three classifiers on KDD'99 and NSL-KDD dataset and Table 6 shows the detection rate of the UNSW-NB15.

### 4.1. Receiver operating characteristic (ROC) curve and area under curve (AUC)

The following Figs. 3–6 depict the ROC and AUC of proposed models. The model RNN has given best performance on outperformed with other proposed models on KDD'99 and NSL-KDD datasets. The model DNN has given good performance on NSL-KDD and UNSW-NB15 datasets. The graphs show the True Positive Rate in comparison to the False Positive Rate. The False Positive Rate is on the x-axis, while the True Positive Rate is on the y-axis. The classification issue performance evaluation at various threshold values is shown by the ROC curve. The likelihood and level of separability are represented by the ROC. The model's performance for detection is improved by a comparatively high AUC. Use of the one vs. rest model may be used to calculate the ROC and AUC values for each class.

### 4.2. Comparative study

The proposed models RNN, LSTM-RNN and DNN implemented to classify the multi attack classification and finding their assessment with classifiers. This process has often been employed in literature. The following figures depict the comparative experiment results. The Fig 7–15 depict the multi attack classification accuracy of models on KDD'99, NSL-KDD and UNSW-NB15 datasets.

The Figs. 7–10 depict the multi attack classification detection rate of models on KDD'99, NSL-KDD and UNSW-NB15 datasets.

The figures to depict the overall performance of proposed models in terms of reliable accuracy, detection rate, precision, recall, F1 score and false positive rate on KDD'99, NSL-KDD and UNSW-NB15 datasets.

From the analysis of the overall performance of RNN, LSTM-RNN and DNN models, we investigated that the models RNN and DNN have performed good accuracy and detection rate on KDD'99 and NSL-KDD dataset. To highlights the robustness of the proposed models, the performance of the model compared to state-of-the-art IDS methodologies, including LSTM-RNN, RBM- DBN, CNN-GRU, Multi-scale CNN, SAE, DBN, DNN, SAE-SVM, S-NDAE, ID-CVAE, SAVAER-DNN, AE and Ensemble methodologies. RNN and DNN models experimental outcomes in terms of accuracy, DR, F1 score and FPR on KDD'99 dataset depict in Table 7 and outcomes on NSL-KDD dataset depict in Table 8.

## 5. Conclusion

The proposed models RNN, LSTM-RNN, and DNN are evaluated on KDD'99, NSL-KDD and UNSW-NB15 datasets. RNN and DNN models are outperformed on KDD'99 and NSL-KDD datasets. The proposed methodologies are unable to provide the considerable results on UNSW-NB15 dataset. From the analysis of the overall performance of RNN, LSTM-RNN and DNN models, we investigated that the models RNN and DNN have performed good accuracy and detection rate on KDD'99 and NSL-KDD dataset. The models F1 score is more comparative accuracy on three datasets. From the literature analysis if F1 score is more comparative accuracy, that states that performance evaluation on imbalanced dataset. The models are not provided the considerable False Positive rate. We used sparse autoencoder (SAE) to balance the dataset. Primarily Sparse autoencoder used to extract the features and reduced dimensionality. The classification algorithms provide significant results on dimensionality reduced datasets.

## Declaration of competing interest

The authors declare that they have no known competing financial interests or personal relationships that could have appeared to influence the work reported in this paper.

## Data availability

No data was used for the research described in the article.